\documentclass[sn-nature,icol]{sn-jnl}% Style for submissions to Nature Portfolio journals
\usepackage{graphicx}%
\usepackage{multirow}%
\usepackage{amsmath,amssymb,amsfonts}%
\usepackage{amsthm}%
\usepackage{mathrsfs}%
\usepackage[title]{appendix}%
\usepackage{xcolor}%
\usepackage{textcomp}%
\usepackage{manyfoot}%
\usepackage{booktabs}%
\usepackage{algorithm}%
\usepackage{algorithmicx}%
\usepackage{algpseudocode}%
\usepackage{listings}%
\usepackage{tikz}

\usepackage{lineno}
\usepackage{geometry}
\usepackage{tabularx}
\usepackage{siunitx}
\DeclareSIUnit \year {yr}
\DeclareSIUnit \erg  {erg}
\DeclareSIUnit \parsec  {pc}
\DeclareSIUnit \gauss  {G}

\theoremstyle{thmstyleone}%
\theoremstyle{thmstyletwo}%

\theoremstyle{thmstylethree}%

\begin{document}

\title[Article Title]{\centering Radiative cooling induced coherent maser emission in relativistic plasmas}

\author[]{\fnm{P.~J.} \sur{Bilbao}}\email{pablojbilbao@tecnico.ulisboa.pt}
\author[]{\fnm{T.} \sur{Silva}}
\author[]{\fnm{L.~O.} \sur{Silva}}\email{luis.silva@tecnico.ulisboa.pt}

\affil[]{\orgdiv{GoLP/Instituto de Plasmas e Fusão Nuclear, Instituto Superior Técnico}, \orgname{Universidade de Lisboa}, \orgaddress{\postcode{1049-001}, \city{Lisboa}, \country{Portugal}}}

\maketitle
\centerline{\large Dated: \today }%, \quad Main body word count: )}

\section*{Abstract}

{\bf Relativistic plasmas in strong electromagnetic fields exhibit distinct properties compared to classical plasmas. In astrophysical environments, such as neutron stars, white dwarfs, AGNs, and shocks, relativistic plasmas are pervasive and are expected to play a crucial role in the dynamics of these systems. Despite their significance, both experimental and theoretical studies of such plasmas have been limited. Here, we present the first ab initio high-resolution kinetic simulations of relativistic plasmas undergoing synchrotron cooling in a highly magnetized medium. Our results demonstrate that these plasmas spontaneously generate coherent linearly polarised radiation (independently of the electron/positron ratio), in a wide range of parameters, via the electron cyclotron maser instability, with radiative losses altering the saturation of this instability. This enables the plasma to continously amplify coherent radiation for significantly longer durations of time. These findings highlight fundamental differences in the behaviour of relativistic plasmas in strongly magnetized environments and align with astronomical phenomena, such as pulsar emission and Fast Radio Bursts.
}
\clearpage
\section*{Main}
Relativistic plasmas are expected to arise around neutron stars, black holes, and other compact objects through various mechanisms, \emph{e.g.}, pair cascades or the Schwinger mechanism \cite{goldreich1969pulsar,timokhin2015polar, philippov2015ab, levinson2018particle,cruz2021coherent}, and in laboratory experiments, \emph{e.g.} with intense lasers or relativistic particle beams \cite{sarri2015generation,grismayer2016laser,chen2023perspectives,arrowsmith2024laboratory, qu2024pair,los2024observation}. These highly energetic plasmas, with electron energies comparable or higher than the rest mass of the particle, form in environments with intense electromagnetic fields, which can sometimes approach the electric and magnetic Schwinger limit ($E_\mathrm{Sc} \simeq 1.3 \times 10^{18}\, \mathrm{V}/\mathrm{m}$ and $B_\mathrm{Sc} \simeq 4.4 \times 10^9\, \mathrm{T}$, respectively) \cite{di2009strong, thomas2012strong, vranic2014all,kaspi2017magnetars, cerutti2017electrodynamics}. 
Under these conditions, quantum electrodynamical (QED) processes, such as non-linear Breit-Wheeler, Compton scattering, and significant radiation reaction, are dominant or comparable to classical plasma processes. Therefore, phenomena such as turbulence \cite{zhdankin2020kinetic,comisso2021pitch,zhou2023magnetogenesis}, shock formation \cite{plotnikov2019synchrotron, vanthieghem2022role}, laser-plasma interactions \cite{di2009strong, thomas2012strong, bulanov2024energy, qu2024pair} beam-plasma interactions \cite{qu2021signature}, and, as shown here, kinetic instabilities, will exhibit significant quantitative differences from their classical plasma counterparts and also manifest qualitatively distinct behaviours and features \cite{uzdensky2019extreme}. 

The complex nature of plasmas in extreme electromagnetic environments has driven significant interest in investigating their kinetic properties. Even simplified electromagnetic field configurations show rich phenomena and can yield surprising results. For instance, plasmas undergoing strong synchrotron cooling have been shown to develop an anisotropic ring shaped momentum distribution, characterised by a population inversion over the Landau levels, $\partial f/\partial p_\perp > 0$, where $f$ represents the plasma momentum distribution and $p_\perp$ the momentum perpendicular to the magnetic field $\mathbf{B}$ \cite{bilbao2022radiation, zhdankin2022synchrotron, bilbao2024ring,ochs2024synchrotron}. However, despite these advances, the collective properties of such strongly magnetised plasmas remain underexplored, particularly regarding the self-consistent electrodynamical effects. Theoretical and experimental investigations are needed to fully understand plasmas in this extreme regime.

One major challenge is that first-principle simulations are constrained by the vast separation of relevant spatial and temporal scales, which differ by several orders of magnitude. This requires the use of massively parallel numerical simulations, which have only recently become feasible at the necessary scales, and with the inclusion all the relevant physics. Replicating these extreme conditions in laboratory experiments is also highly challenging, and only recent advancements in experimental techniques have made it possible to generate plasmas under such extreme environments \cite{chen2023perspectives,arrowsmith2024laboratory, los2024observation}.

In this study, we address this gap by conducting the first and largest scale, first-principles numerical simulations that demonstrate how relativistic plasmas embedded in strong magnetic fields can spontaneously produce linearly polarized coherent radiation via the electron cyclotron maser instability (ECMI) \cite{sprangle1977linear,chen1991unified,bingham2000generation,melrose2017coherent}. We demonstrate how the instability is qualitatively modified by the inclusion of synchrotron losses.
This phenomenon occurs in collisionless plasmas with relativistic temperatures, where radiation reaction plays a crucial role. Our simulations reveal that synchrotron cooling first establishes the Landau population inversion, in the shape of a ring momentum distribution, and then maintains it for longer timescales than previously thought possible, leading to continuous coherent amplification of radiation and a modified saturation state of the maser instability.

\subsection*{Results}

\begin{figure*}[]
    \centering
    \begin{tikzpicture}
        \draw (0, 0) node[inner sep=0] {\includegraphics[width=0.328\textwidth]{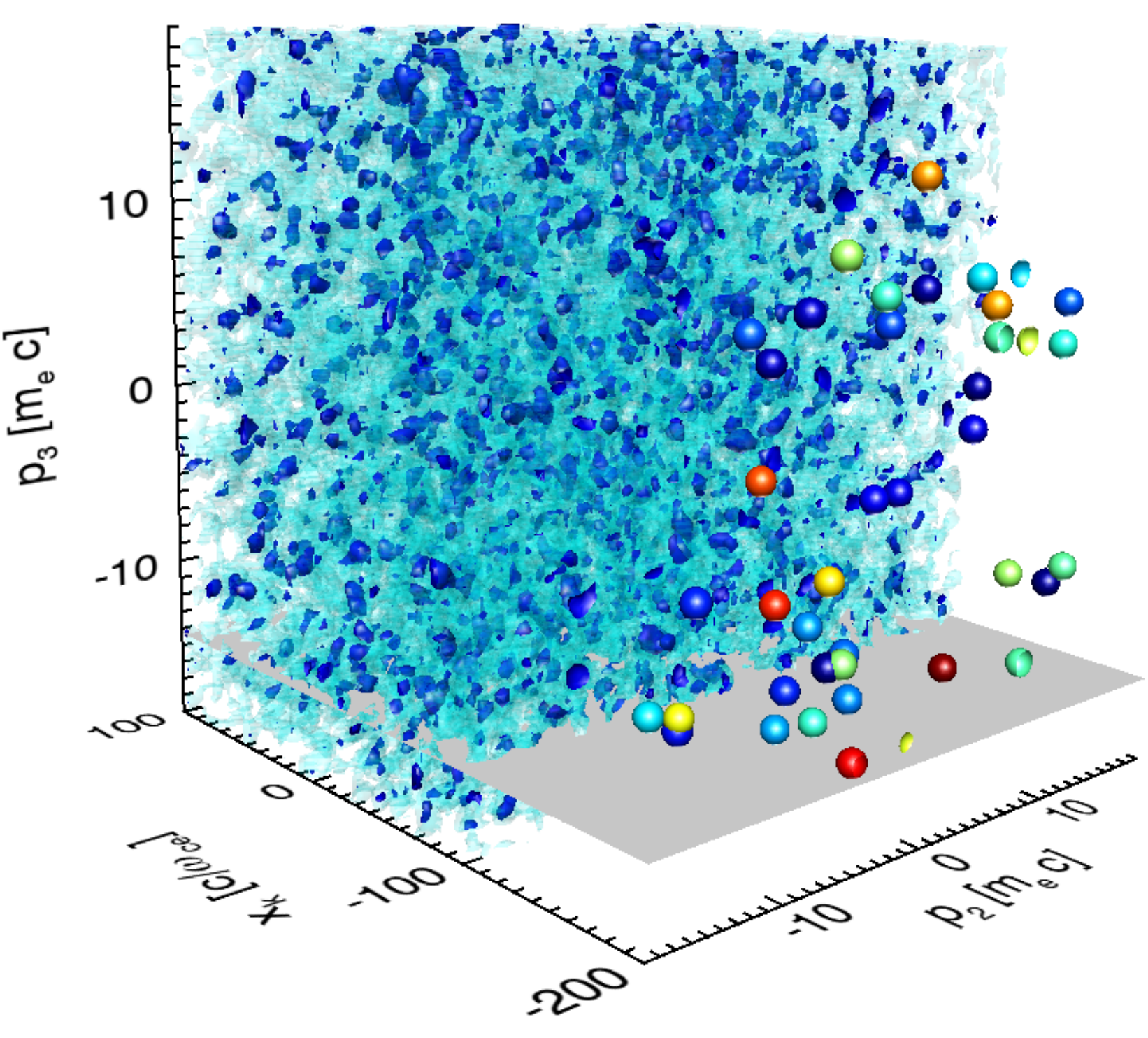}};
        \draw (0, 2.6) node {\small{$t=0000\, [\omega_{ce}^{-1}]$}};
    \end{tikzpicture}
    \begin{tikzpicture}
        \draw (0, 0) node[inner sep=0] {\includegraphics[width=0.328\textwidth]{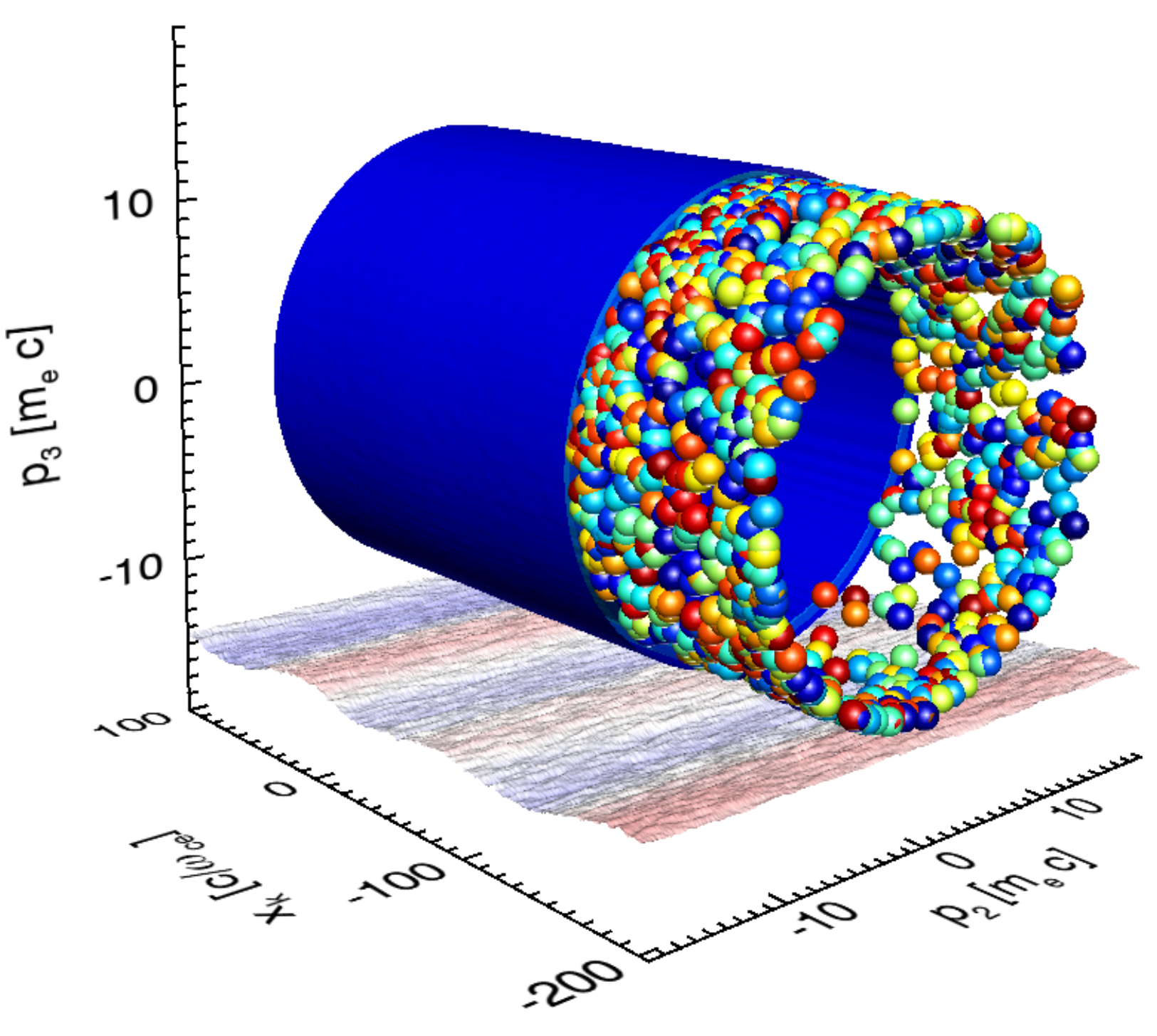}};
        \draw (0, 2.6) node {\small{$t=7350\, [\omega_{ce}^{-1}]$}};
    \end{tikzpicture}
    \begin{tikzpicture}
        \draw (0, 0) node[inner sep=0] {\includegraphics[width=0.328\textwidth]{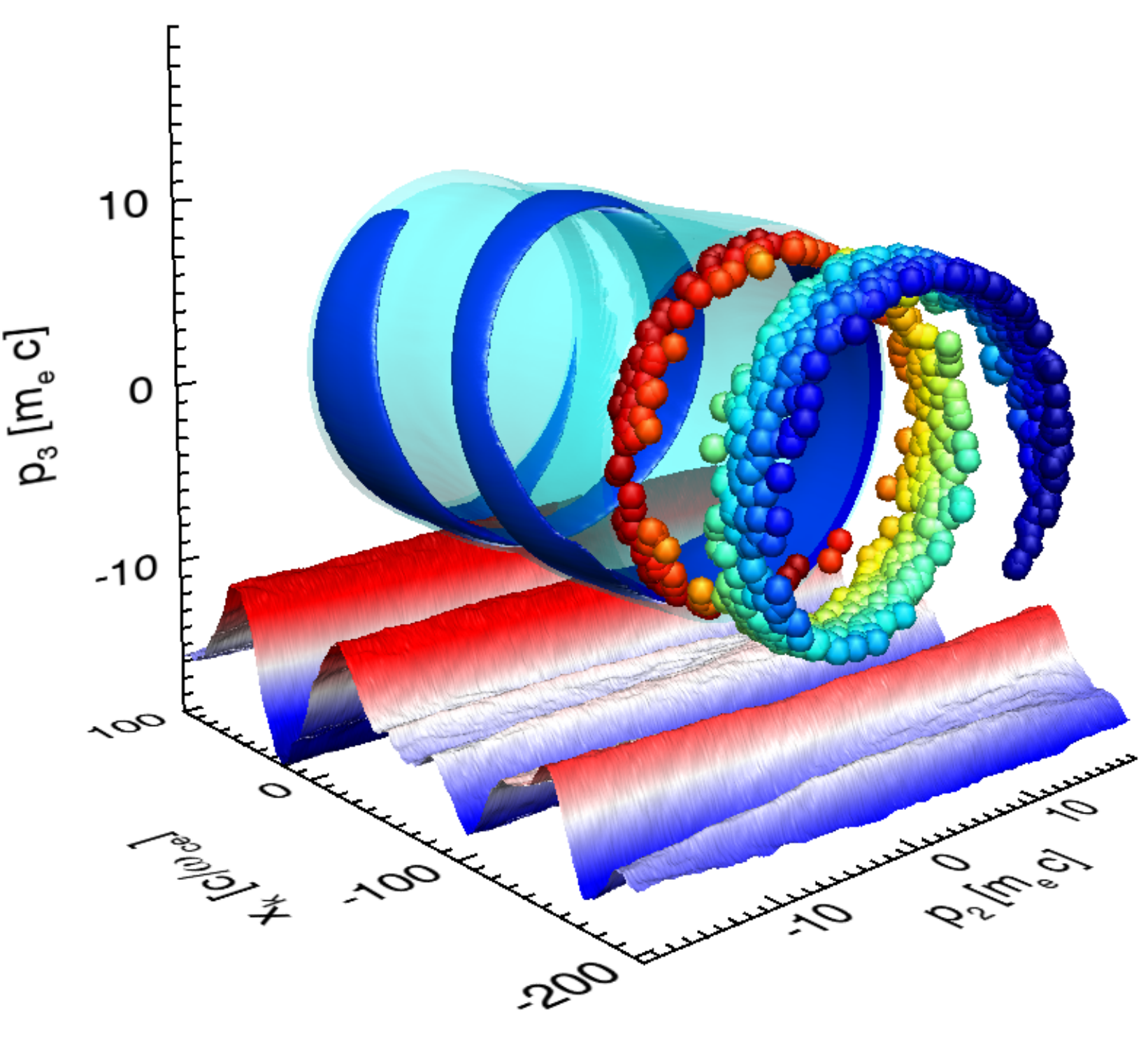}};
        \draw (0, 2.6) node {\small{$t=9200\, [\omega_{ce}^{-1}]$}};
    \end{tikzpicture}
    \caption{Self-consistent particle-in-cell simulations demonstrate the onset of electron cyclotron maser instability (ECMI) and coherent amplification of radiation. The temporal evolution (from left to right) of the plasma distribution function $f(\mathbf{r}, \mathbf{p}, t)$, shown as a 3D projection, $p_2$-$p_3$-$x_k$, of the 6D phase-space, where $p_2$ and $p_3$ are the momentum directions perpendicular to the $B$-field, with $B/B_\mathrm{Sc} = 0.002$, and $x_k$ is the spatial direction along the propagation of the X mode (also perpendicular to $B$). Two isosurfaces of the distribution function are represented: the light blue and blue surfaces are at 0.5 and 0.8 of the peak value of the distribution function, respectively. The projection in the bottom plane (red-blue colours) represents the electric field associated with the amplified electromagnetic wave (X-mode). A sample of the plasma particles is also shown, with the colour representing their azimuthal phase (between $0$ and $2 \pi$, from red to blue) with respect to the amplified electromagnetic wave Left: Initially, the plasma consists of a Maxwellian thermal plasma with initial momentum distribution $f_0\propto e^{-\left|\mathbf{p}\right|^2/(2 p_{th}^2)}$, where $p_{th}=1000\, m_e c$, and no X-mode is observed, the particles have random phases; Middle: a ring momentum distribution function has developed due to synchrotron cooling and amplification of the X-mode begins, but particles are still arranged in random phases; Right: the amplified X-mode is evident and the ring is azimuthally bunched along the direction of propagation of the X-mode, as seen by the spiral structure in phase space, and the clear phase alignment, as seen by the colour of the particles, following the corresponding colour scale. }\label{fig:ring_onset}
\end{figure*}
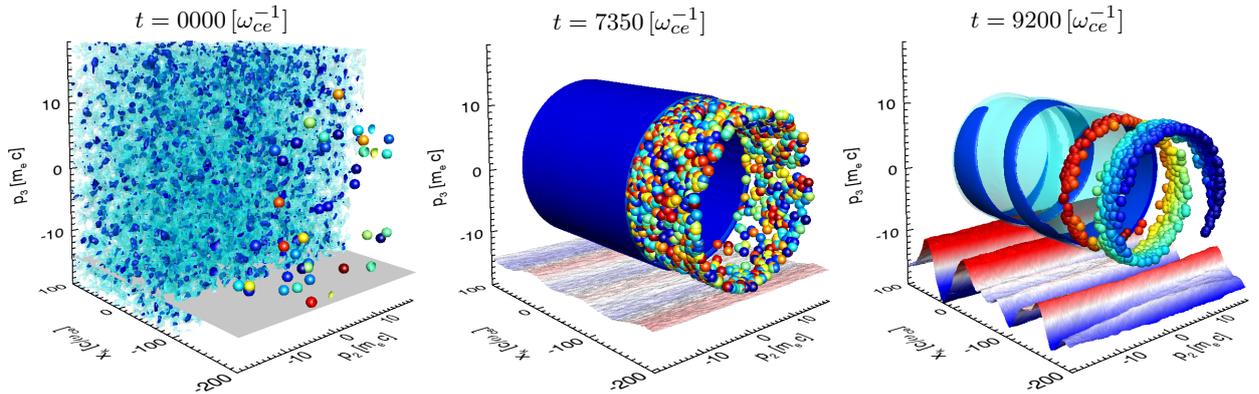
We have investigated, through large-scale, high-resolution particle-in-cell (PIC) simulations (see Methods), tenuous pair plasmas embedded in strong magnetic fields, where the cyclotron frequency $\omega_{ce} = eB/m_e$ is much larger than the plasma frequency $\omega_{pe}= \sqrt{n_e e^2/(\varepsilon_0 m_e)}$ where $e$ and $m_e$ are the electron charge and electron mass, respectively; $n_e$ is the pair plasma density, $\varepsilon_0$ is the permittivity of free space. To the best of our knowledge, these simulations represent the most extensive study in terms of both spatial and temporal scales for this system to date (see Methods section). PIC simulation results, shown in Fig. \ref{fig:ring_onset}, illustrate that an initial thermal plasma, in its proper rest frame, evolves into a ring momentum distribution characterised by steep gradients in the perpendicular momentum component $p_\perp$ with respect to the $B$ field, and a narrow energy spread $\Delta p_\perp$. This evolution triggers the efficient onset of the electron cyclotron maser instability, which coherently amplifies electromagnetic thermal fluctuations in the magnetised plasma, generating X-mode electromagnetic waves. This process leads to azimuthal bunching, a characteristic signature of the instability, as confirmed by the phase-space projection after the onset of the instability, shown in Fig. \ref{fig:ring_onset}.

The evolution of the synchrotron-cooled plasma, the subsequent growth rate of the X-mode, and the dynamical timescales of the instability, as seen in the simulations, can be directly explained from kinetic theory with the inclusion of radiative losses \cite{landau1975classical, kuz1978bogolyubov, bilbao2022radiation, zhdankin2022synchrotron,bilbao2024ring}. For a tenuous pair-plasma in a strong magnetic field, the momentum distribution of a plasma evolves as $f(p_\perp, p_\parallel, \tau ) = f\left( \frac{p_\perp}{1-\tau p_\perp}, \frac{p_\parallel}{1-\tau p_\perp}\right)/\left( 1-\tau p_\perp\right)^4$, where $p_\perp$ is normalised to $m_e c$, $\tau$ is a normalised time such that $\tau = \frac{2\alpha}{3} B_0 \omega_{ce} t $, where $B_0 = B/B_\mathrm{Sc}$ and $\alpha$ is the fine-structure constant \cite{bilbao2022radiation}. An important feature is that $f$ is bounded between $0 < p_\perp < 1/(2\alpha B_0 \omega_{ce}t)$ and that the resulting ring radius asymptotically approaches the boundary at $p_\perp = 3/(2\alpha B_0 \omega_{ce}t)$ \cite{bilbao2024ring}. Thus, a relativistic plasma, independently of the initial shape of $f$, will develop into an anisotropic ring momentum distribution \cite{bilbao2022radiation, zhdankin2022synchrotron,bilbao2024ring, ochs2024synchrotron}, which will fulfil the conditions for efficient maser emission, as the simulations here demonstrate. 

The growth rate of X-modes, $\Gamma$, as a function of $t$ and the wave angular frequency $\omega$ can be approximated by \cite{alexandrov1984principles}
\begin{equation}
    \Gamma(\omega, t) = 2\pi^2 \frac{\omega_{pe}^2}{\omega }\sum_{n=1}^{\infty} \left\{ p_\perp'^2  \left.\frac{\partial f_\perp(t)}{\partial p_\perp}\right|_{p_\perp = p_\perp'} \left[J_n'\left( \frac{\omega p_\perp'}{n \omega_{ce}}\right)\right]^2\right\}, \label{eq:growthrate_lim}
\end{equation}
where $J_n'(b)$ is the first derivative of the $n$th order Bessel function evaluated at $b$, $p_\perp' = \sqrt{n^2\omega_{ce}^2 /\omega^2 - 1}$ is the resonant momentum with the given frequency $\omega$, $f_\perp(p_\perp,t)$ is the perpendicular momentum distribution, \emph{i.e.}, integrated along the magnetic field direction $f_\perp(p_\perp,t) = \int_{-\infty}^\infty f(p_\perp, p_\parallel, t) \partial p_\parallel$, where it is assumed that the time dependence of the distribution function is much slower than $\omega$, analogous to the WKB approximation. Equation (\ref{eq:growthrate_lim}) shows some of the key universal emission features characteristic of the synchrotron-induced electron cyclotron maser: i) the amplification rate is proportional to the plasma density $\omega_{pe}^2 \propto n_e$, ii) it occurs near the different cyclotron harmonic resonances $\omega_{ce}/\gamma$, and iii) is proportional to $\partial f_\perp / \partial p_\perp$. 

As the distribution function cools down it develops a small region between the ring radius and the edge of momentum space, of width $\Delta p \simeq 3\sqrt{3}/(2\alpha B_0 p_{th}\omega_{ce} t)^2$, where $p_{th}$ is the initial thermal spread. The width of the ring allows to estimate the maximum growth rate at the fundamental frequency. By approximating $\Gamma_\mathrm{max}\simeq 18 \pi \frac{\omega_{pe}^2 }{\omega_{ce}} \frac{p_R \gamma_R}{\Delta p^2 } \left[J_1'(1)\right]^2 \simeq  2 \pi\alpha^2 B_0^2 p_{th}^2 \omega_{pe}^2 \omega_{ce} t^2$, $\Gamma_\mathrm{max}$ shows how an initial plasma with a higher thermal spread $p_{th}$ leads to a higher growth rate, and how the growth rate increases with time due to the distribution $f$ developing a stepper gradient, as the plasma cools.

\begin{figure}[]
    \centering
    \includegraphics[width=0.5\textwidth]{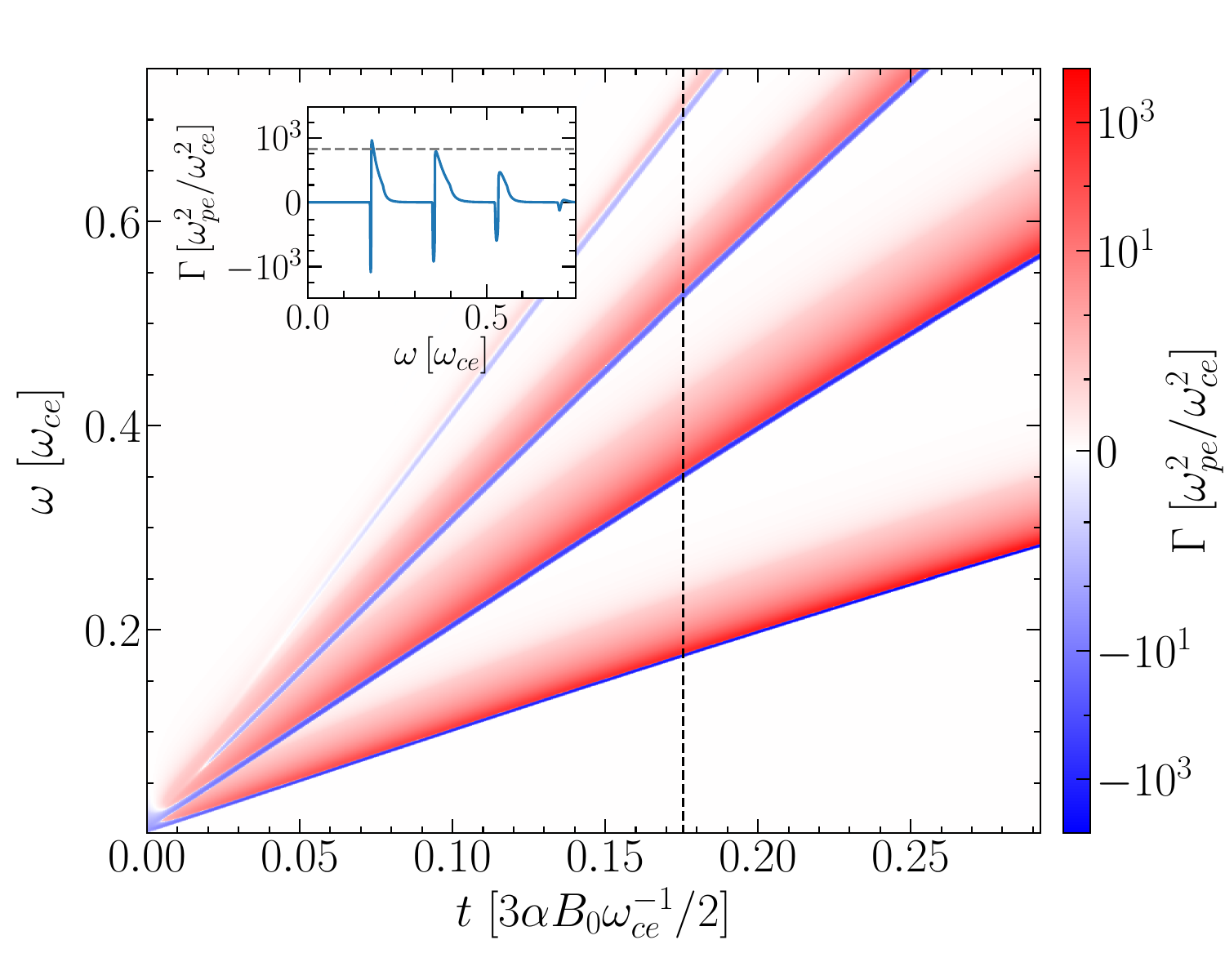}
    \caption{The key emission features of the cyclotron maser instability are demonstrated by temporal evolution of the X-mode growth rate, $\Gamma(\omega, t)$, as described by Eq. (\ref{eq:growthrate_lim}), for a typical distribution function, $f_\perp(p_\perp, t=0) = e^{-p_\perp^2/(2 p_{th}^2)}/p_{th}^3 (2\pi)^{1/2}$, where $p_{th} = 100\ m_e c$. Since the ring distribution is a general characteristic of hot plasmas (\emph{i.e.}, those with $p_{th} \gg m_e c$) undergoing synchrotron cooling, this initial distribution function effectively represents the qualitative behaviour of maser emission under various initial conditions \cite{bilbao2022radiation,bilbao2024ring}. The figure demonstrates that the emission is evenly spaced in $\omega$-space, as evident from the line-out of $\Gamma(\omega, t=0.165\tau)$ shown in the inset, which highlights the emission and absorption regions, where $\Gamma>0$ and $\Gamma<0$, respectively. The emission predominantly occurs near the harmonics of the resonant frequency, which gradually converge towards the harmonics of $\omega_{ce}$ as the ring distribution cools down and asymptotically approaches $p_\perp=0$.}
    \label{fig:growthrate}
\end{figure}

To illustrate the properties of the ECMI with radiative cooling, we determine $\Gamma$ (shown in Fig. \ref{fig:growthrate}) for an initial Maxwellian distribution (this greatly simplifies analytical calculations). The growth rate and dynamics of the ECMI do not significantly vary with the initial distribution shape (\emph{e.g.} Maxwell-J\"uttner), as plasmas with large energy spreads will always converge to a ring distribution \cite{bilbao2022radiation, bilbao2024ring}. Simulations with an initial Maxwell-Jüttner distribution (not shown here) confirm the same ECMI dynamics as those with an initial Maxwellian distribution.

The synchrotron cooled plasma produces a small emission region in $\omega$ above the frequency $n\omega_{ce}/\gamma_r$ and a small absorption region below that frequency, for each harmonic $n$, as observed in Fig. \ref{fig:growthrate}. The emission changes with time due to the shifting resonance condition as the ring cools down, and the estimate for $\Gamma_\mathrm{max}$ is confirmed (c.f. inset axis of Fig. \ref{fig:growthrate}).

For the efficient onset of the ECMI two condtions must be met: i) that $\Gamma>0$, and ii) that the ring distribution provides several e-foldings to the interacting wave, before dephasing. Initially, the ring forms and develops $\partial f/\partial p_\perp>0$, within a small region of momentum space of width $\Delta p \simeq 3\sqrt{3}/(2\alpha B_0 p_{th}^2 \omega_{ce} t)^2$ below the ring radius. At that point, waves with $\omega$ in resonance within that region of momentum space are amplified at a rate comparable to $\Gamma_\mathrm{max}$. 
This amplification continues for a short time interval $\Delta t_d=3\sqrt{3}/(2 \alpha B_0 p_\mathrm{th}^2 \omega_{ce})$ until the ring cools down enough that it dephases and $\omega$ is no longer in resonance. Now the ring is in resonance with a slightly higher $\omega$, and the process repeats itself. 

Initially, the cooling rate is too rapid to efficiently amplify any frequency before the ring dephases, \emph{i.e.} $\Delta t_d<\Gamma_{max}^{-1}$. As $\Gamma_\mathrm{max}$ increases over time, eventually, $\Gamma_{max}^{-1}$ becomes smaller than $\Delta t_d$, allowing the ring to remain in resonance for several e-folding times and provide efficient amplification, which occurs, in the proper frame of the plasma, at time 
\begin{equation}
    t_o = \left(\frac{1}{\sqrt{3} \pi \alpha}\right)^{1/2} \frac{eB_\mathrm{Sc} /m_e}{\omega_{pe}}\frac{B_0^{1/2}}{ p_{th}^{1/2}}  \omega_{ce}^{-1} ,  \label{eq:onset_time}
\end{equation}
or $t_o\, [\SI{12}{\micro\second}] \simeq(B\, [\SI{1}{\mega\gauss}]\, n_e\, [10^6\,\mathrm{cm}^{-3}]\ p_{th}\, [100\,m_e c])^{-1/2}$; $t_o$ determines the time at which coherent emission begins, from the beginning of the cooling process. The scaling of Eq. (\ref{eq:onset_time}) has been confirmed with PIC simulations (see Supplemental Material).

\begin{figure}[]
\centering
    \includegraphics[width=0.5\textwidth]{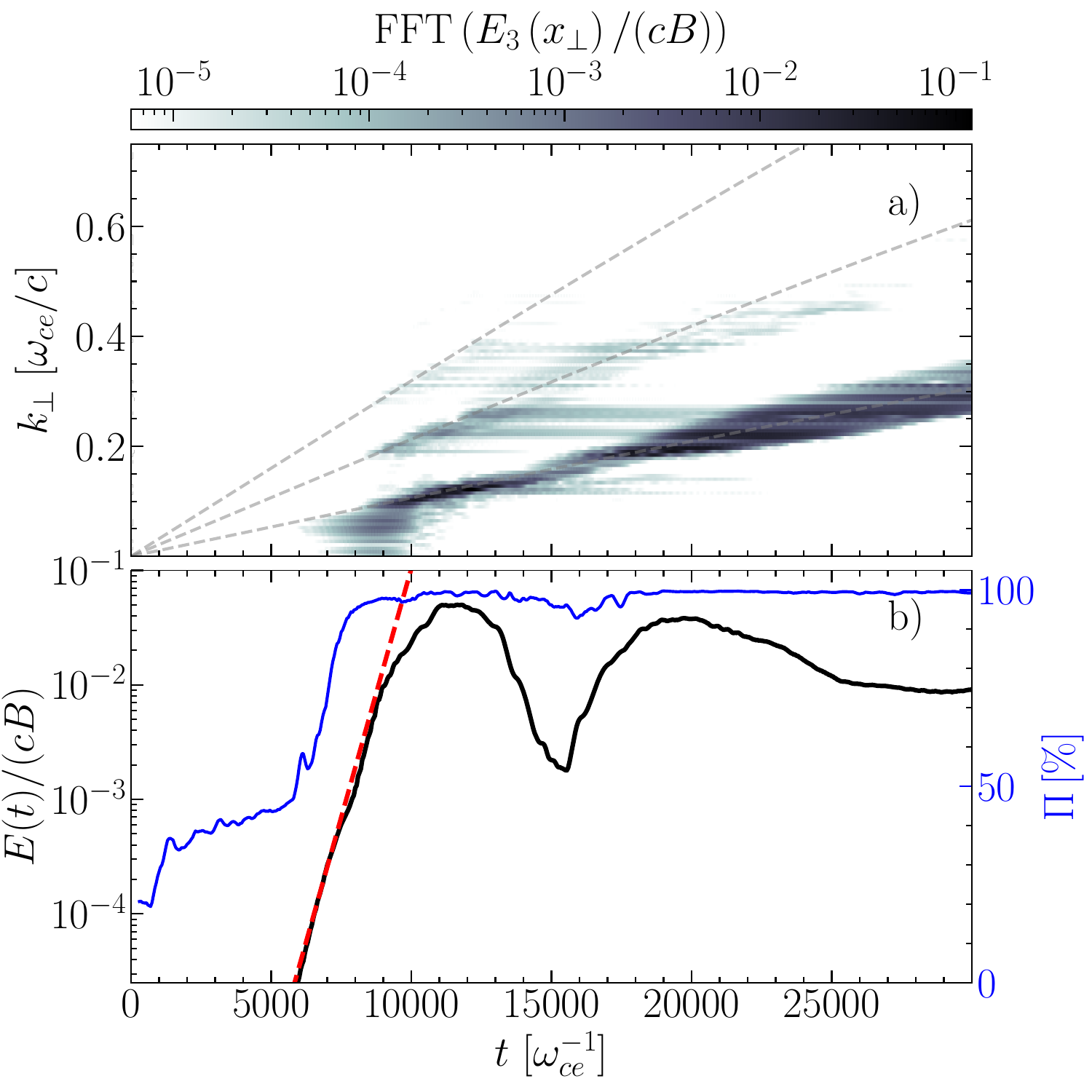}
    \caption{Particle-in-cell simulations illustrate the temporal evolution of the X-mode electromagnetic spectrum and energy, which agree with the theoretical estimates, both qualitative and quantitatively. Panel a) shows the Fourier transform of the electric field component $E_3$ which is perpendicular to both the ambient magnetic field (of amplitude $B$) and the wave vector $\mathbf{k}$. This spectrum shows the ongoing evolution of the electromagnetic fields during both the linear and non-linear phases of the instability. Theoretical predictions for the maximum growth rates (i.e., the resonant frequencies) from Eq. (\ref{eq:growthrate_lim}) are overlaid as grey dashed lines. Panel b) shows the normalised time evolution of the electric field amplitude (black line) and the degree of polarization, derived from the Stokes parameters (blue line). The estimate for the linear growth rate $\Gamma$, obtained by numerically solving the full dispersion relation, is compared with the simulation results after the onset of the maser instability but still at early times ($t_o<t\lesssim12000\,\omega_{ce}^{-1}$), shown as a red dashed line, which follows the relation $E_3\propto e^{\Gamma t}$.}\label{fig:fft_evo}
\end{figure}

The electromagnetic spectrum of the amplified X-mode is shown in Figure \ref{fig:fft_evo}.a, and demonstrates that the X-mode spectrum peaks near the regions with the highest $\Gamma$, with the first three harmonics being amplified, whereas the 4th is much weaker, as analytically predicted (see Fig. \ref{fig:growthrate}). The degree of polarisation, in Fig. \ref{fig:growthrate}.b, demonstrates that the self-consistent radiation resulting from emission is highly linearly polarised. This is a result of the X-mode being the fastest-growing mode in highly magnetized plasmas \cite{winglee1985fundamental}.

The classical ECMI should reach a steady state for the radiated spectrum at saturation. The saturation of the amplified wave occurs due to phase trapping. 
Before saturation, and for the electrons to provide energy to the wave, it is necessary that $\delta\omega=\omega - \omega_{ce}/\gamma\gtrsim0$ \cite{sprangle1977linear}. At saturation ($t\sim 1.2 \times10^{4}\, \omega_{ce}^{-1}$ in Fig. \ref{fig:fft_evo}), particles overshoot the phase trapping condition and now $\delta\omega\lesssim0$: particles extract energy from the previously amplified wave. This is a well-known phenomenon \cite{sprangle1977linear}, that explains the dip in electric field amplitude at $t\sim 1.5 \times10^{4}\, \omega_{ce}^{-1}$ in Fig. \ref{fig:fft_evo}.b and the saturated state, as shown in devices such as Gyrotrons, FELs, and ICLs \cite{sprangle1977linear, chen1991unified}. PIC simulations capture the ECMI well beyond the linear regime and demonstrate that synchrotron-cooled plasmas evolve differently in the non-linear stage of the instability, as seen in Fig. \ref{fig:fft_evo}.a: surprisingly, the spectrum continues to evolve with the X-mode spectrum shifting to higher frequencies and widening the spectrum. The emitted radiation is trapped in our simulation domain, mimicking an infinite plasma volume. For a finite volume plasma the X-modes escape and convert to light waves, our numerical simulations with finite plasmas confirm this picture (not shown here).

\begin{figure}[]
    \centering
    \includegraphics[width=0.5\textwidth]{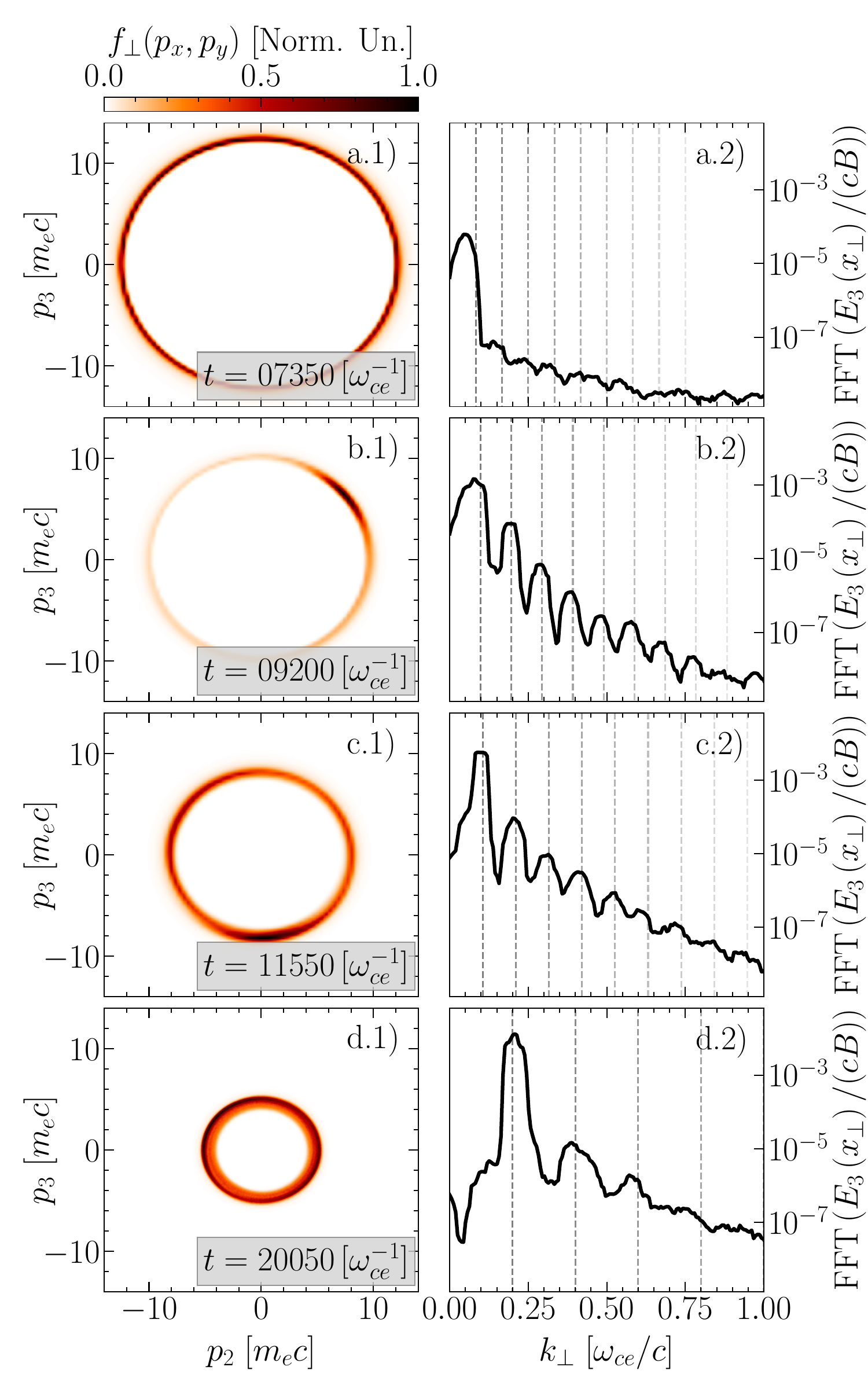}
    \caption{
    PIC simulations illustrate that the ring distribution is sustained beyond the linear regime of the instability. This is seen in the evolution of a radiatively cooled electron-positron plasma undergoing the electron cyclotron maser instability, starting from an initial Maxwellian distribution with a thermal momentum of $p_\mathrm{th}=1000\,m_ec$. Column 1 displays the electron momentum distribution integrated along the magnetic field direction $\mathbf{x_1}$, $f_\perp \left( p_2, p_3\right)$, while Column 2 shows the spectrum of the associated X-mode electromagnetic wave, with dashed lines indicating the harmonics in resonance with the ring radius, as predicted by our theoretical model. Each row (a-d) corresponds to different times in the simulation, with time progressing from top to bottom: a) shows the developed ring distribution from an initial Maxwellian plasma; with b) shows the onset of the electron cyclotron maser instability, characterised by phase trapping (the positrons, not shown here, bunch up on the opposite phase of the ring), typical of the transition from the linear to the non-linear phase of this instability, with narrow emission observed in the spectrum. c) shows the widening of the ring distribution and a corresponding shift in the emission as the system reaches the standard ECMI saturation point. d) demonstrates the further evolution of the distribution as the instability transitions to the non-linear regime, where the synchrotron cooling competes with the ECMI, causing the ring to widen and the emission to broaden accordingly.
    }\label{fig:rings_evo}
\end{figure}
The evolution of the emission spectrum is explained by a competition between the maser instability, which diffuses the population inversion via phase-bunching, and the synchrotron losses, which reinforce the population inversion.
The evolution of the perpendicular momentum distribution of the plasma, as it transitions from the linear to the non-linear regime of the instability, confirms this competition. After the establishment of the ring distribution, with a narrow momentum spread (Fig. \ref{fig:rings_evo}.a.1), the onset of the instability produces the azimuthal bunching characteristic of the instability (Fig. \ref{fig:rings_evo}.b.1) and the efficient amplification of the different harmonics  (Fig. \ref{fig:rings_evo}.b.2). At saturation, the ring has continued to constrict, as seen by the smaller radius (Fig. \ref{fig:rings_evo}.c.1) and the frequency of each harmonic has been upshifted slightly (Fig. \ref{fig:rings_evo}.c.2). Due to the interaction with the wave the ring attains a wider momentum spread and a flatter gradient of the distribution function with respect to perpendicular momentum $\partial f_\perp/\partial p_\perp$. In classical plasmas, the instability would quenched, leading to the end of the amplification as the classical ECMI saturates.

Unlike the classical ECMI, in the synchrotron-cooled instability, as electrons continue to cool, then $\partial f/\partial p_\perp$ increases at lower $p_\perp$, thus, enhancing the amplification of higher-frequency waves resonant with the ring, resulting in a smooth upshift of the electromagnetic wave (Fig.  \ref{fig:fft_evo}.a and Fig. \ref{fig:rings_evo}). Additionally, the amplification of the wave causes the ring distribution to widen through phase trapping. This widening is counterbalanced by the bunching effect provided by radiative cooling (Fig. \ref{fig:rings_evo}.d.1). Consequently, the EM spectrum broadens at late times, indicating that the saturated stage does not inhibit population inversion. We have observed that the ring structure in momentum space is not completely diffused in the amplification process and the ECMI is active well beyond the times presented in this work.

Eventually, the ring will cool down to $T\sim m_e c^2 /\sqrt{3}$, and the instability overcomes the bunching process (as the synchrotron losses which sustain the ring become less important \cite{bilbao2024ring}) and the ring is diffused. This occurs when the ring cools down to a ring radius $p_R \sim m_e c/\sqrt{3}$, which determines for how long the ring structure and emission can be sustained. In the plasma proper frame that is comparable to $t_\mathrm{em} \simeq 2 \omega_{ce}^{-1}/\sqrt{3}\alpha B_0$, or $t_\mathrm{em} \left[\SI{400}{\micro \second}\right] \simeq \left(B\, \left[\SI{1}{\mega\gauss}\right]\right)^{-2}$, which is independent of the plasma parameters, as it is a timescale determined solely by the cooling process when $t_{em}> t_o $. 

Therefore, after the onset of the maser at $t_o$ (Eq. \ref{eq:onset_time}) the ring and the ECMI, will be sustained until $t_{em}$, producing a long pulse of radiation.
This finding addresses a major criticism of the ECMI as a source of ``long-lived" coherent radiation \cite{lu2018radiation}. In relativistic plasmas, the onset of the maser instability does not disrupt the population inversion, allowing continued emission. Radiative effects sustain the population inversion, enabling the maser to operate over prolonged periods. Interestingly, depending on the plasma parameters, the ring can form and cool below $p_R \sim m_e c/\sqrt{3}$, before the onset of the ECMI, and in that case $t_o>t_\mathrm{em}$. In this scenario, the onset of the ECMI can still happen, and the ring will begin to diffuse right after saturation. Radiative losses will not be able to reform or sustain the ring, resulting in a single pulse of electromagnetic waves that escapes the plasma, resembling the classical ECMI emission.

\subsection*{Discussion}
\begin{figure*}
    \centering
    \includegraphics[width=\textwidth]{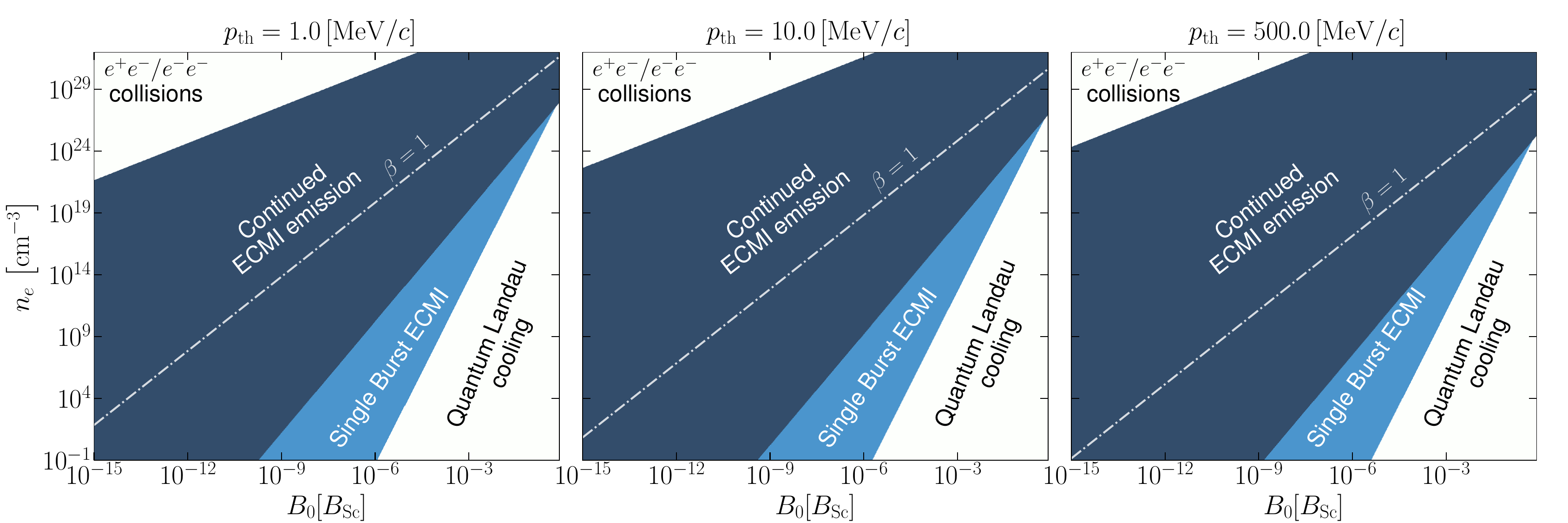}
    \caption{The comparison of relevant timescales for the onset of the synchrotron ECMI demonstrates that maser emission occurs for a wide range of plasma and magnetic field parameters. As presented across three panels, each corresponding to different initial thermal energies and defined by plasma density, magnetic field strength, and initial thermal energy. Regions in dark blue indicate where the maser instability triggers and sustains continued emission, occurring before competing processes like collisional relaxation or cooling to quantum Landau levels can interfere. The top left white region highlights conditions where collisional relaxation dominates, diffusing the ring structure and preventing efficient maser emission, while the right-hand white region shows where cooling to quantum Landau levels occurs before the maser instability can develop, rendering the plasma degenerate and preventing maser onset. A light blue region represents conditions where the maser instability triggers with a single burst produced, as the onset occurs after the ring has cooled below the threshold thermal momentum needed for sustained bunching. A white dashed line marks the boundary where the plasma beta parameter ($\beta=1$), the ratio of plasma pressure to magnetic field pressure, is reached. Above this line, other instabilities, such as the synchrotron firehose instability can also trigger.}
    \label{fig:time}
\end{figure*}
The necessary plasma parameters for relativistic plasmas to emit coherent radiation via ECMI can be determined by guaranteeing that a hierarchy of timescales is fulfilled. Firstly, the onset of the instability $t_o$ (Eq. \ref{eq:onset_time}) must be shorter than any diffusive process, \emph{e.g.} Coulomb collisions, Compton collisions and pair annihilation (See Supplemental Material). Additionally, the onset time must be shorter than $t_\mathrm{landau}\simeq\frac{2}{3\alpha} B_0^{-2}\omega_{ce}^{-1}$, \emph{i.e.}, the time it takes the ring to cool to the the lowest quantum Landau level when the plasma becomes degenerate.

There is a wide range of parameters for which the synchrotron-induced ECMI operates efficiently before (Coulomb or Compton) collisional relaxation and cooling to the quantum levels can take place, see Fig. \ref{fig:time}. The parameter space increases as the plasma becomes more energetic. The ECMI will spontaneously trigger in a wide range of plasma parameters of relevance to astrophysical emissions and observations \cite{melrose2017coherent}.
Other instabilities can also trigger due to synchrotron losses; the firehose instability can operate when $\beta>1$, where $\beta$ is the ratio of the plasma pressure to the magnetic field pressure \cite{zhdankin2022synchrotron}. In that regime, the firehose instability will modify the momentum distribution, but as the plasma cools down the plasma can transition to a $\beta<1$ regime enabling a modified version of the ``ring'' ECMI to operate. Preliminary PIC simulations have confirmed this picture and will be explored elsewhere.

The fastest-growing mode, the linearly polarized X-mode, efficiently converts to light waves as it exits the strongly magnetized plasma, achieving nearly 100\% efficiency in the transition from X-mode to light mode (for $\omega_{ce} \gg \omega_{pe}$) \cite{stix1992waves}. This emission spans a wide frequency range, depending on the magnetic field parameters, and includes radio frequencies. In the proper frame of the plasma, the emitted frequencies occur at harmonics of $n \omega_{ce}/\gamma_r$, where $\gamma_r = (1+p_r^2/m_e^2c^2)^{1/2}$ is the Lorentz factor associated with the ring radius at the onset of the maser instability. Radio emission occurs in the proper frame at $\omega\, [\SI{17}{\tera\hertz}]\simeq B\, [\SI{1}{\mega \gauss}] / \gamma_r\, [1]$. Additionally, the emission maintains a constant ratio of $\Delta \omega/\omega$, where $\Delta \omega$ is the frequency separation between different harmonics. This constant ratio is observed in sources such as the Crab pulsar \cite{hankins2015crab}.

%Due to time dilation and the relativistic Doppler effect, the received radiation is upshifted by a factor of $\gamma_b$, where $\gamma_b$ is the Lorentz factor of the beam's proper frame. Furthermore, the emission timescale becomes shorter by a factor of $1/\gamma_b$, and the intensity increases by a factor of $\gamma_b$ due to relativistic beaming \cite{lind1985semidynamical}.

%\section*{Conclusions}
Using the largest PIC simulations to date for tenuous, synchrotron-cooled plasmas, we have demonstrated that these plasmas can spontaneously produce coherent radiation when self-consistent electrodynamical radiative effects are considered. This radiation is driven by the onset of the ring-driven electron cyclotron maser instability. Importantly, our results reveal that this emission process can persist for significantly longer periods than previously thought, due to the interplay between the instability and synchrotron losses. This finding challenges the classical understanding of ECMI, which has traditionally been seen as resulting in only short bursts of radiation due to rapid saturation, and demonstrates the relevance of ECMI in synchrotron cooled relativistic plasmas.

Our findings suggests the synchrotron driven ECMI is relevant beyond the specific plasma conditions explored in our simulations, having also a broader applicability across various plasma and magnetic field parameters. This implies that this mechanism can operate in a wide range of astrophysical environments. Notably, this mechanism appears to align with several key features observed in pulsar and magnetar emissions \cite{philippov2022pulsar}, offering a potential explanation for certain characteristics of Fast Radio Bursts and pulsar emission. The observed connection between FRBs and magnetars, particularly the detection of FRBs coinciding with magnetar glitches, suggests a model where pair-plasmas in a low-twist magnetar magnetosphere generate these bursts \cite{timokhin2015polar,bochenek2020fast,chime2022sub}. Our findings demonstrate that synchrotron losses combined with the ECMI can sustain coherent emission over longer durations, which may explain the observed coherence, radio emission range, linear polarisation, repetition, and similarities across diverse astronomical objects. This mechanism is a result of the new qualitative properties of extreme plasmas. The resulting synchrotron-induced ECMI mechanism has implications that extend beyond the specific case of FRBs to a broader spectrum of astrophysical phenomena and future astro-laboratory experiments.

\bibliography{main.bib}

\providecommand{\noopsort}[1]{}\providecommand{\singleletter}[1]{#1}%\providecommand{\noopsort}[1]{}\providecommand{\singleletter}[1]{#1}%
\begin{thebibliography}{10}
\expandafter\ifx\csname url\endcsname\relax
  \def\url#1{\burl{#1}}\fi
\expandafter\ifx\csname urlprefix\endcsname\relax\def\urlprefix{URL }\fi
\providecommand{\bibinfo}[2]{#2}
\providecommand{\eprint}[2][]{\url{#2}}
\providecommand{\doi}[1]{\url{https://doi.org/#1}}
\bibcommenthead

\bibitem{goldreich1969pulsar}
\bibinfo{author}{Goldreich, P.} \& \bibinfo{author}{Julian, W.~H.}
\newblock \bibinfo{title}{Pulsar electrodynamics}.
\newblock \emph{\bibinfo{journal}{Astrophysical Journal, vol. 157, p. 869}} \textbf{\bibinfo{volume}{157}}, \bibinfo{pages}{869} (\bibinfo{year}{1969}).

\bibitem{timokhin2015polar}
\bibinfo{author}{Timokhin, A.} \& \bibinfo{author}{Harding, A.}
\newblock \bibinfo{title}{On the polar cap cascade pair multiplicity of young pulsars}.
\newblock \emph{\bibinfo{journal}{The Astrophysical Journal}} \textbf{\bibinfo{volume}{810}}, \bibinfo{pages}{144} (\bibinfo{year}{2015}).

\bibitem{philippov2015ab}
\bibinfo{author}{Philippov, A.~A.}, \bibinfo{author}{Spitkovsky, A.} \& \bibinfo{author}{Cerutti, B.}
\newblock \bibinfo{title}{Ab initio pulsar magnetosphere: three-dimensional particle-in-cell simulations of oblique pulsars}.
\newblock \emph{\bibinfo{journal}{The Astrophysical Journal Letters}} \textbf{\bibinfo{volume}{801}}, \bibinfo{pages}{L19} (\bibinfo{year}{2015}).

\bibitem{levinson2018particle}
\bibinfo{author}{Levinson, A.} \& \bibinfo{author}{Cerutti, B.}
\newblock \bibinfo{title}{Particle-in-cell simulations of pair discharges in a starved magnetosphere of a kerr black hole}.
\newblock \emph{\bibinfo{journal}{Astronomy \& Astrophysics}} \textbf{\bibinfo{volume}{616}}, \bibinfo{pages}{A184} (\bibinfo{year}{2018}).

\bibitem{cruz2021coherent}
\bibinfo{author}{Cruz, F.}, \bibinfo{author}{Grismayer, T.}, \bibinfo{author}{Chen, A.~Y.}, \bibinfo{author}{Spitkovsky, A.} \& \bibinfo{author}{Silva, L.~O.}
\newblock \bibinfo{title}{Coherent emission from qed cascades in pulsar polar caps}.
\newblock \emph{\bibinfo{journal}{The Astrophysical Journal Letters}} \textbf{\bibinfo{volume}{919}}, \bibinfo{pages}{L4} (\bibinfo{year}{2021}).

\bibitem{sarri2015generation}
\bibinfo{author}{Sarri, G.} \emph{et~al.}
\newblock \bibinfo{title}{Generation of neutral and high-density electron--positron pair plasmas in the laboratory}.
\newblock \emph{\bibinfo{journal}{Nature communications}} \textbf{\bibinfo{volume}{6}}, \bibinfo{pages}{6747} (\bibinfo{year}{2015}).

\bibitem{grismayer2016laser}
\bibinfo{author}{Grismayer, T.}, \bibinfo{author}{Vranic, M.}, \bibinfo{author}{Martins, J.~L.}, \bibinfo{author}{Fonseca, R.} \& \bibinfo{author}{Silva, L.}
\newblock \bibinfo{title}{Laser absorption via quantum electrodynamics cascades in counter propagating laser pulses}.
\newblock \emph{\bibinfo{journal}{Physics of Plasmas}} \textbf{\bibinfo{volume}{23}}, \bibinfo{pages}{056706} (\bibinfo{year}{2016}).

\bibitem{chen2023perspectives}
\bibinfo{author}{Chen, H.} \& \bibinfo{author}{Fiuza, F.}
\newblock \bibinfo{title}{Perspectives on relativistic electron--positron pair plasma experiments of astrophysical relevance using high-power lasers}.
\newblock \emph{\bibinfo{journal}{Physics of Plasmas}} \textbf{\bibinfo{volume}{30}} (\bibinfo{year}{2023}).

\bibitem{arrowsmith2024laboratory}
\bibinfo{author}{Arrowsmith, C.} \emph{et~al.}
\newblock \bibinfo{title}{Laboratory realization of relativistic pair-plasma beams}.
\newblock \emph{\bibinfo{journal}{Nature Communications}} \textbf{\bibinfo{volume}{15}}, \bibinfo{pages}{5029} (\bibinfo{year}{2024}).

\bibitem{qu2024pair}
\bibinfo{author}{Qu, K.}, \bibinfo{author}{Griffith, A.} \& \bibinfo{author}{Fisch, N.~J.}
\newblock \bibinfo{title}{Pair filamentation and laser scattering in beam-driven qed cascades}.
\newblock \emph{\bibinfo{journal}{Physical Review E}} \textbf{\bibinfo{volume}{109}}, \bibinfo{pages}{035208} (\bibinfo{year}{2024}).

\bibitem{los2024observation}
\bibinfo{author}{Los, E.} \emph{et~al.}
\newblock \bibinfo{title}{Observation of quantum effects on radiation reaction in strong fields}.
\newblock \emph{\bibinfo{journal}{arXiv preprint arXiv:2407.12071}}  (\bibinfo{year}{2024}).

\bibitem{di2009strong}
\bibinfo{author}{Di~Piazza, A.}, \bibinfo{author}{Hatsagortsyan, K.} \& \bibinfo{author}{Keitel, C.}
\newblock \bibinfo{title}{Strong signatures of radiation reaction below the radiation-dominated regime}.
\newblock \emph{\bibinfo{journal}{Physical Review Letters}} \textbf{\bibinfo{volume}{102}}, \bibinfo{pages}{254802} (\bibinfo{year}{2009}).

\bibitem{thomas2012strong}
\bibinfo{author}{Thomas, A.}, \bibinfo{author}{Ridgers, C.}, \bibinfo{author}{Bulanov, S.}, \bibinfo{author}{Griffin, B.} \& \bibinfo{author}{Mangles, S.}
\newblock \bibinfo{title}{Strong radiation-damping effects in a gamma-ray source generated by the interaction of a high-intensity laser with a wakefield-accelerated electron beam}.
\newblock \emph{\bibinfo{journal}{Physical Review X}} \textbf{\bibinfo{volume}{2}}, \bibinfo{pages}{041004} (\bibinfo{year}{2012}).

\bibitem{vranic2014all}
\bibinfo{author}{Vranic, M.}, \bibinfo{author}{Martins, J.~L.}, \bibinfo{author}{Vieira, J.}, \bibinfo{author}{Fonseca, R.~A.} \& \bibinfo{author}{Silva, L.~O.}
\newblock \bibinfo{title}{All-optical radiation reaction at 1 0 21 w/cm 2}.
\newblock \emph{\bibinfo{journal}{Physical Review Letters}} \textbf{\bibinfo{volume}{113}}, \bibinfo{pages}{134801} (\bibinfo{year}{2014}).

\bibitem{kaspi2017magnetars}
\bibinfo{author}{Kaspi, V.~M.} \& \bibinfo{author}{Beloborodov, A.~M.}
\newblock \bibinfo{title}{Magnetars}.
\newblock \emph{\bibinfo{journal}{Annual Review of Astronomy and Astrophysics}} \textbf{\bibinfo{volume}{55}}, \bibinfo{pages}{261--301} (\bibinfo{year}{2017}).

\bibitem{cerutti2017electrodynamics}
\bibinfo{author}{Cerutti, B.} \& \bibinfo{author}{Beloborodov, A.~M.}
\newblock \bibinfo{title}{Electrodynamics of pulsar magnetospheres}.
\newblock \emph{\bibinfo{journal}{Space Science Reviews}} \textbf{\bibinfo{volume}{207}}, \bibinfo{pages}{111--136} (\bibinfo{year}{2017}).

\bibitem{zhdankin2020kinetic}
\bibinfo{author}{Zhdankin, V.}, \bibinfo{author}{Uzdensky, D.~A.}, \bibinfo{author}{Werner, G.~R.} \& \bibinfo{author}{Begelman, M.~C.}
\newblock \bibinfo{title}{Kinetic turbulence in shining pair plasma: intermittent beaming and thermalization by radiative cooling}.
\newblock \emph{\bibinfo{journal}{Monthly Notices of the Royal Astronomical Society}} \textbf{\bibinfo{volume}{493}}, \bibinfo{pages}{603--626} (\bibinfo{year}{2020}).

\bibitem{comisso2021pitch}
\bibinfo{author}{Comisso, L.} \& \bibinfo{author}{Sironi, L.}
\newblock \bibinfo{title}{Pitch-angle anisotropy controls particle acceleration and cooling in radiative relativistic plasma turbulence}.
\newblock \emph{\bibinfo{journal}{Physical Review Letters}} \textbf{\bibinfo{volume}{127}}, \bibinfo{pages}{255102} (\bibinfo{year}{2021}).

\bibitem{zhou2023magnetogenesis}
\bibinfo{author}{Zhou, M.}, \bibinfo{author}{Zhdankin, V.}, \bibinfo{author}{Kunz, M.~W.}, \bibinfo{author}{Loureiro, N.~F.} \& \bibinfo{author}{Uzdensky, D.~A.}
\newblock \bibinfo{title}{Magnetogenesis in a collisionless plasma: from weibel instability to turbulent dynamo}.
\newblock \emph{\bibinfo{journal}{The Astrophysical Journal}} \textbf{\bibinfo{volume}{960}}, \bibinfo{pages}{12} (\bibinfo{year}{2023}).

\bibitem{plotnikov2019synchrotron}
\bibinfo{author}{Plotnikov, I.} \& \bibinfo{author}{Sironi, L.}
\newblock \bibinfo{title}{The synchrotron maser emission from relativistic shocks in fast radio bursts: 1d pic simulations of cold pair plasmas}.
\newblock \emph{\bibinfo{journal}{Monthly Notices of the Royal Astronomical Society}} \textbf{\bibinfo{volume}{485}}, \bibinfo{pages}{3816--3833} (\bibinfo{year}{2019}).

\bibitem{vanthieghem2022role}
\bibinfo{author}{Vanthieghem, A.} \emph{et~al.}
\newblock \bibinfo{title}{The role of plasma instabilities in relativistic radiation-mediated shocks: stability analysis and particle-in-cell simulations}.
\newblock \emph{\bibinfo{journal}{Monthly Notices of the Royal Astronomical Society}} \textbf{\bibinfo{volume}{511}}, \bibinfo{pages}{3034--3045} (\bibinfo{year}{2022}).

\bibitem{bulanov2024energy}
\bibinfo{author}{Bulanov, S.} \emph{et~al.}
\newblock \bibinfo{title}{On the energy spectrum evolution of electrons undergoing radiation cooling}.
\newblock \emph{\bibinfo{journal}{Fundamental Plasma Physics}} \bibinfo{pages}{100036} (\bibinfo{year}{2024}).

\bibitem{qu2021signature}
\bibinfo{author}{Qu, K.}, \bibinfo{author}{Meuren, S.} \& \bibinfo{author}{Fisch, N.~J.}
\newblock \bibinfo{title}{Signature of collective plasma effects in beam-driven qed cascades}.
\newblock \emph{\bibinfo{journal}{Physical review letters}} \textbf{\bibinfo{volume}{127}}, \bibinfo{pages}{095001} (\bibinfo{year}{2021}).

\bibitem{uzdensky2019extreme}
\bibinfo{author}{Uzdensky, D.} \emph{et~al.}
\newblock \bibinfo{title}{Extreme plasma astrophysics}.
\newblock \emph{\bibinfo{journal}{arXiv preprint arXiv:1903.05328}}  (\bibinfo{year}{2019}).

\bibitem{bilbao2022radiation}
\bibinfo{author}{Bilbao, P.~J.} \& \bibinfo{author}{Silva, L.~O.}
\newblock \bibinfo{title}{Radiation reaction cooling as a source of anisotropic momentum distributions with inverted populations}.
\newblock \emph{\bibinfo{journal}{Physical Review Letters}} \textbf{\bibinfo{volume}{130}}, \bibinfo{pages}{165101} (\bibinfo{year}{2023}).

\bibitem{zhdankin2022synchrotron}
\bibinfo{author}{Zhdankin, V.}, \bibinfo{author}{Kunz, M.~W.} \& \bibinfo{author}{Uzdensky, D.~A.}
\newblock \bibinfo{title}{Synchrotron firehose instability}.
\newblock \emph{\bibinfo{journal}{The Astrophysical Journal}} \textbf{\bibinfo{volume}{944}}, \bibinfo{pages}{24} (\bibinfo{year}{2023}).

\bibitem{bilbao2024ring}
\bibinfo{author}{Bilbao, P.~J.}, \bibinfo{author}{Ewart, R.~J.}, \bibinfo{author}{Assun{\c{c}}ao, F.}, \bibinfo{author}{Silva, T.} \& \bibinfo{author}{Silva, L.~O.}
\newblock \bibinfo{title}{Ring momentum distributions as a general feature of vlasov dynamics in the synchrotron dominated regime}.
\newblock \emph{\bibinfo{journal}{Physics of Plasmas}} \textbf{\bibinfo{volume}{31}} (\bibinfo{year}{2024}).

\bibitem{ochs2024synchrotron}
\bibinfo{author}{Ochs, I.~E.}
\newblock \bibinfo{title}{Synchrotron-driven instabilities in relativistic plasmas of arbitrary opacity}.
\newblock \emph{\bibinfo{journal}{arXiv preprint arXiv:2407.13106}}  (\bibinfo{year}{2024}).

\bibitem{sprangle1977linear}
\bibinfo{author}{Sprangle, P.} \& \bibinfo{author}{Drobot, A.}
\newblock \bibinfo{title}{The linear and self-consistent nonlinear theory of the electron cyclotron maser instability}.
\newblock \emph{\bibinfo{journal}{IEEE Transactions on Microwave Theory and Techniques}} \textbf{\bibinfo{volume}{25}}, \bibinfo{pages}{528--544} (\bibinfo{year}{1977}).

\bibitem{chen1991unified}
\bibinfo{author}{Chen, K.-R.}, \bibinfo{author}{Dawson, J.}, \bibinfo{author}{Lin, A.} \& \bibinfo{author}{Katsouleas, T.}
\newblock \bibinfo{title}{Unified theory and comparative study of cyclotron masers, ion-channel lasers, and free electron lasers}.
\newblock \emph{\bibinfo{journal}{Physics of Fluids B: Plasma Physics}} \textbf{\bibinfo{volume}{3}}, \bibinfo{pages}{1270--1278} (\bibinfo{year}{1991}).

\bibitem{bingham2000generation}
\bibinfo{author}{Bingham, R.} \& \bibinfo{author}{Cairns, R.}
\newblock \bibinfo{title}{Generation of auroral kilometric radiation by electron horseshoe distributions}.
\newblock \emph{\bibinfo{journal}{Physics of Plasmas}} \textbf{\bibinfo{volume}{7}}, \bibinfo{pages}{3089--3092} (\bibinfo{year}{2000}).

\bibitem{melrose2017coherent}
\bibinfo{author}{Melrose, D.}
\newblock \bibinfo{title}{Coherent emission mechanisms in astrophysical plasmas}.
\newblock \emph{\bibinfo{journal}{Reviews of Modern Plasma Physics}} \textbf{\bibinfo{volume}{1}}, \bibinfo{pages}{1--81} (\bibinfo{year}{2017}).

\bibitem{landau1975classical}
\bibinfo{author}{Landau, L.~D.} \& \bibinfo{author}{Lifshitz, E.~M.}
\newblock \emph{\bibinfo{title}{The Classical Theory of Fields}} Vol.~\bibinfo{volume}{2} (\bibinfo{publisher}{Pergamon Press, Oxford}, \bibinfo{year}{1975}).

\bibitem{kuz1978bogolyubov}
\bibinfo{author}{Kuz'menkov, L.}
\newblock \bibinfo{title}{The bogolyubov hierarchy of equations for relativistic systems. radiation damping of waves in a plasma} \textbf{\bibinfo{volume}{23}}, \bibinfo{pages}{469--471} (\bibinfo{year}{1978}).

\bibitem{alexandrov1984principles}
\bibinfo{author}{Alexandrov, A.~F.}, \bibinfo{author}{Bogdankevich, L.~S.}, \bibinfo{author}{Rukhadze, A.~A.} \emph{et~al.}
\newblock \emph{\bibinfo{title}{Principles of plasma electrodynamics}} Vol.~\bibinfo{volume}{9} (\bibinfo{publisher}{Springer}, \bibinfo{year}{1984}).

\bibitem{winglee1985fundamental}
\bibinfo{author}{Winglee, R.}
\newblock \bibinfo{title}{Fundamental and harmonic electron cyclotron maser emission}.
\newblock \emph{\bibinfo{journal}{Journal of Geophysical Research: Space Physics}} \textbf{\bibinfo{volume}{90}}, \bibinfo{pages}{9663--9674} (\bibinfo{year}{1985}).

\bibitem{lu2018radiation}
\bibinfo{author}{Lu, W.} \& \bibinfo{author}{Kumar, P.}
\newblock \bibinfo{title}{On the radiation mechanism of repeating fast radio bursts}.
\newblock \emph{\bibinfo{journal}{Monthly Notices of the Royal Astronomical Society}} \textbf{\bibinfo{volume}{477}}, \bibinfo{pages}{2470--2493} (\bibinfo{year}{2018}).

\bibitem{stix1992waves}
\bibinfo{author}{Stix, T.~H.}
\newblock \emph{\bibinfo{title}{Waves in plasmas}}  (\bibinfo{publisher}{Springer Science \& Business Media}, \bibinfo{year}{1992}).

\bibitem{hankins2015crab}
\bibinfo{author}{Hankins, T.}, \bibinfo{author}{Jones, G.} \& \bibinfo{author}{Eilek, J.}
\newblock \bibinfo{title}{The crab pulsar at centimeter wavelengths. i. ensemble characteristics}.
\newblock \emph{\bibinfo{journal}{The Astrophysical Journal}} \textbf{\bibinfo{volume}{802}}, \bibinfo{pages}{130} (\bibinfo{year}{2015}).

\bibitem{philippov2022pulsar}
\bibinfo{author}{Philippov, A.} \& \bibinfo{author}{Kramer, M.}
\newblock \bibinfo{title}{Pulsar magnetospheres and their radiation}.
\newblock \emph{\bibinfo{journal}{Annual Review of Astronomy and Astrophysics}} \textbf{\bibinfo{volume}{60}}, \bibinfo{pages}{495--558} (\bibinfo{year}{2022}).

\bibitem{bochenek2020fast}
\bibinfo{author}{Bochenek, C.~D.} \emph{et~al.}
\newblock \bibinfo{title}{A fast radio burst associated with a galactic magnetar}.
\newblock \emph{\bibinfo{journal}{Nature}} \textbf{\bibinfo{volume}{587}}, \bibinfo{pages}{59--62} (\bibinfo{year}{2020}).

\bibitem{chime2022sub}
\bibinfo{author}{{CHIME/FRB Collaboration}} \emph{et~al.}
\newblock \bibinfo{title}{Sub-second periodicity in a fast radio burst}.
\newblock \emph{\bibinfo{journal}{Nature}} \textbf{\bibinfo{volume}{607}}, \bibinfo{pages}{256--259} (\bibinfo{year}{2022}).

\bibitem{fonseca2002osiris}
\bibinfo{author}{Fonseca, R.~A.} \emph{et~al.}
\newblock \emph{\bibinfo{title}{Osiris: A three-dimensional, fully relativistic particle in cell code for modeling plasma based accelerators}}, \bibinfo{pages}{342--351} (\bibinfo{organization}{Springer}, \bibinfo{year}{2002}).

\bibitem{vranic2016classical_m}
\bibinfo{author}{Vranic, M.}, \bibinfo{author}{Martins, J.~L.}, \bibinfo{author}{Fonseca, R.~A.} \& \bibinfo{author}{Silva, L.~O.}
\newblock \bibinfo{title}{Classical radiation reaction in particle-in-cell simulations}.
\newblock \emph{\bibinfo{journal}{Computer Physics Communications}} \textbf{\bibinfo{volume}{204}}, \bibinfo{pages}{141--151} (\bibinfo{year}{2016}).

\bibitem{vranic2016quantum_m}
\bibinfo{author}{Vranic, M.}, \bibinfo{author}{Grismayer, T.}, \bibinfo{author}{Fonseca, R.~A.} \& \bibinfo{author}{Silva, L.~O.}
\newblock \bibinfo{title}{Quantum radiation reaction in head-on laser-electron beam interaction}.
\newblock \emph{\bibinfo{journal}{New Journal of Physics}} \textbf{\bibinfo{volume}{18}}, \bibinfo{pages}{073035} (\bibinfo{year}{2016}).

\bibitem{rybicki1991radiative}
\bibinfo{author}{Rybicki, G.~B.} \& \bibinfo{author}{Lightman, A.~P.}
\newblock \emph{\bibinfo{title}{Radiative processes in astrophysics}}  (\bibinfo{publisher}{John Wiley \& Sons}, \bibinfo{year}{1991}).

\bibitem{trubnikov1965particle}
\bibinfo{author}{Trubnikov, B.}
\newblock \bibinfo{title}{Particle interactions in a fully ionized plasma}.
\newblock \emph{\bibinfo{journal}{Rev. Plasma Phys.}} \textbf{\bibinfo{volume}{1}} (\bibinfo{year}{1965}).

\bibitem{goldston1995introduction}
\bibinfo{author}{Goldston, R.} \& \bibinfo{author}{Rutherford, P.}
\newblock \emph{\bibinfo{title}{Introduction to plasma physics}}  (\bibinfo{publisher}{IOP Publishing, UK}, \bibinfo{year}{1995}).

\bibitem{lightman1982relativistic}
\bibinfo{author}{Lightman, A.}
\newblock \bibinfo{title}{Relativistic thermal plasmas-pair processes and equilibria}.
\newblock \emph{\bibinfo{journal}{The Astrophysical Journal}} \textbf{\bibinfo{volume}{253}}, \bibinfo{pages}{842--858} (\bibinfo{year}{1982}).

\bibitem{jauch2012theory}
\bibinfo{author}{Jauch, J.~M.} \& \bibinfo{author}{Rohrlich, F.}
\newblock \emph{\bibinfo{title}{The theory of photons and electrons: the relativistic quantum field theory of charged particles with spin one-half}}  (\bibinfo{publisher}{Springer Science \& Business Media}, \bibinfo{year}{2012}).

\end{thebibliography}
%\clearpage

\section*{Methods}
\subsection*{Particle-in-cell simulations}
We study radiatively cooled rings and the subsequent ECMI via particle-in-cell (PIC) simulations with OSIRIS \cite{fonseca2002osiris}, including classical \cite{vranic2016classical_m} and QED \cite{vranic2016quantum_m} radiation reaction.
The PIC method is widely used in plasma physics to model the behaviour of plasmas by solving the equations of motion for charged particles and the self-consistent evolution of electromagnetic fields. The plasma is represented by a large number of particles, which move according to the Lorentz force (with Landau-Lifshiftz force to account for semi-classical radiative losses \cite{vranic2016classical_m, landau1975classical}, and QED Monte Carlo module to account for QED processes \cite{vranic2016quantum_m}) in response to the electromagnetic fields. These fields, in turn, are computed on a grid using Maxwell's equations, with the particle motions and fields updated iteratively.

PIC simulations are well-suited for studying kinetic instabilities in plasmas, such as the electron cyclotron maser instability (ECMI), because they capture the full range of particle interactions and non-linear effects. The massively parallel nature of these simulations allows for the handling of large-scale problems, making it possible to explore complex phenomena in tenuous, synchrotron-cooled plasmas with high fidelity and at unprecedented scales.

For the simulations presented in this work, we have considered the setup described analytically. The ECMI is a kinetic instability for which the relevant dynamics occur in momentum space, and the resulting excited wave modes propagate either parallel or perpendicular to the magnetic field. Therefore, our simulations employ a 2D configuration space and full 3D momentum space. This guarantees that all the relevant physics of the ECMI and cooling dynamics are captured in our setup. There is a magnetic field aligned along the $x_1$-direction of strength $B = 100\ \textrm{GigaGauss}$, \emph{i.e.} $B_0 \simeq 0.002$ normalized to the Schwinger field ($B_{Sc} = 4.4 \times 10 ^9 \ \textrm{Teslas}$). The magnetic field has a cyclotron frequency $\omega_{ce}=|e|B/m_e = 1.75\times 10^{18}\ \textrm{Hz}$. All relevant timescales and lengths are normalized to the cyclotron period $\omega_{ce}^{-1}$ and $c/\omega_{ce}$. The simulations utilize a small timestep such that the cyclotron period is accurately resolved $\Delta t = 0.014 \omega_{ce}^{-1}$ and a spatial resolution of $\Delta x = 0.02 c/\omega_{ce}$ (in both directions), which fulfils the 2D Courant condition $\Delta x > 2^{-1/2} c\Delta t$. We resolve the gyromotion of all electrons with at least $\sim70$ points. The simulation window has a size of $L_1 = 200\ c/\omega_{ce}$ and $L_2 = 1000 c\omega_{ce}$, this yields a simulation grid of size $10000\times50000$ and we employ $16$ particles per cell, \emph{i.e.} a total of 16 billion computational particles. In turn, resolving the gyroradius of mildly relativistic electrons with at least $50$ points, with higher energy particles being resolved with even more spatial and temporal points, due to the Larmor radius $r_L \propto \gamma$.

The dimensions of the simulation box were carefully chosen, the $x_2$-direction utilises a larger domain to capture the theoretically predicted modes that propagate perpendicular and almost perpendicular to the magnetic field (a 1D3V PIC simulation would solely capture the the wave dynamics propagating perfectly parallel to the simulation domain, but not those at small angles).

The simulations were run for $40000 \omega_{ce}^{-1}$, this is $\sim2.8\times 10 ^{6}$ time iterations. Due to the size of each simulation, the high-resolution runs were performed in LUMI-C (Finland) and had an average cost of 1 million CPU hours. Smaller simulations were performed in Deucalion (Portugal).

The plasma was initialized with a plasma frequency ratio of $\omega_{pe}/\omega_{ce} \simeq 0.00223$ for each species. This value was carefully chosen to create a low-density electron-positron pair plasma within the simulation domain, employing periodic boundary conditions. The simulation encompasses several skin depths and Debye lengths throughout the cooling process and the development of the instability, effectively modelling a tenuous pair plasma in a strong magnetic field where $\omega_{pe}/\omega_{ce} \ll 1$. The pair plasma is initialised from a Maxwellian momentum distribution $f\left( p_\perp, p_\parallel \right)\propto e^{-\left(p_\perp^2 + p_\parallel^2\right)/\left( 2 p_{th}^2\right)}$, with $p_{th}= 1000\ m_e c$. Simulations with a Maxwell-J\"uttner distribution were also performed and, as expected, provided the same numerical results. 

The macro-particles employ cubic interpolation. Different current smoothing filters were tested, and we found that first-order binomial smoothing was sufficient to reduce the computational collisionality, for the large number of time steps in the simulations. For the simulations shown in this work, OSIRIS employed the Landau-Lifshitz model for classical radiation reaction as described by \cite{vranic2016classical_m}. Moreover, QED simulations which employ a Monte Carlo method to model quantum synchrotron emission \cite{vranic2016quantum_m}, were also used, QED simulations agree with simulations employing the Landau-Lifshitz pusher with the inclusion of stochastic diffusion, as expected as $\chi = p B_0/(m_e c)$ decreases rapidly during the cooling process. These effects will be explored elsewhere at higher energies, where diffusive effects are expected to be dominant.

Convergence studies were performed to determine the computational parameters and to ensure energy conservation accounting for synchrotron losses.

It is important to note that all results, both numerical and analytical, are presented in the proper frame of the plasma or beam. This means that the results can be directly applied to beam-plasma systems in other reference frames through the appropriate Lorentz transformation.

\subsection*{Degree of polarization of the maser radiation}
The Stokes parameters and degree of polarization can be obtained from the PIC simulations presented in this work. For an observer with a line of sight along $\mathbf{\hat{x_2}}$, perpendicular to the magnetic field along $\mathbf{\hat{x_1}}$. The Stokes parameters in the Jones basis are given by \cite{rybicki1991radiative}
\begin{align}
    I &= \left<E_1^2\right>+\left<E_3^2\right>\\
    Q &= \left<E_3^2\right>-\left<E_3^2\right>\\
    U &= \left<E_a^2\right>-\left<E_b^2\right>\\
    V &= \left<E_r^2\right>-\left<E_l^2\right>,
\end{align}
where the subscripts $a$, $b$ represent the electric field components projected onto the cartesian basis rotated 45º. The subscripts $l$ and $r$ is the projection in the circular basis such that $\mathbf{\hat{l}} = (\mathbf{\hat{x_1}} +i\mathbf{\hat{x_3}})/\sqrt{2}$ and $\mathbf{\hat{r}} = (\mathbf{\hat{x_1}} -i\mathbf{\hat{x_3}})/\sqrt{2}$. The $\left<E_i^2\right>$ represents the averaged quantity such that $\left<E_i^2\right> = \frac{1}{T}\int_0^T E_i^2(t) dt$, where $E_i^2(t)$ is the square of the electric field at a given time at the observer's position. From the simulation results we can construct the different quantities to in their respective Jones basis and averaging them spatially to synthetically obtain the Stokes parameters an observer would measure.

The degree of linear polarization is then defined as \cite{rybicki1991radiative}
\begin{equation}
    \Pi = \sqrt{Q^2 + U^2}/I.\label{supeq:sup:Pi}
\end{equation}
The result from this synthetic diagnostic for the percentile for the degree of polarization, in the PIC simulations, is shown in Fig. \ref{fig:fft_evo}.b. Initially, the electromagnetic radiation is mostly unpolarized, later once the onset of the electron cyclotron maser instability ($t\sim6000\, \omega_{ce}^{-1}$) the radiation becomes highly linearly polarized reaching a maximum $\Pi=99.8\%$.

\bmhead{Acknowledgments}
We thank Prof. R. Bingham, Prof. A. R. Bell, Mr. D. Carvalho, Mr. R. J. Ewart, Prof. M Lyutikov, Dr. B. Malaca, Dr. R. Torres, Dr. T. Grismayer, and Dr. V. Zhdankin, for their useful discussions. This work was supported by FCT (Portugal) (Grant UI/BD/151559/2021 and X-MASER - 2022.02230.PTDC). Simulations were performed at LUMI (Finland) funded by EuroHPC-JU project EHPC-REG-2021R0038 and at Deucalion (Portugal) funded by FCT Masers in Astrophysical Plasmas (MAPs) I.P project 2024.11062.CPCA.A3.

\bmhead{Author contributions}
All authors contributed significantly to this work.

\bmhead{Competing interests}
The authors declare no competing interests.

\bmhead{Supplementary information}
Supplementary Information file provided.

\backmatter
\clearpage

%\clearpage

%\section*{Tables}

%\section*{Figures}

\clearpage

\renewcommand{\figurename}{Supplementary Fig.}
\renewcommand{\tablename}{Supplementary Table}
\renewcommand*\contentsname{Supplementary information}
\setcounter{figure}{0}

\section*{\centering{\large  Supplemental Material: Radiative cooling induced coherent maser emission in relativistic plasmas}}

\subsection*{Parameter scan of the onset time of the ECMI}
\begin{figure*}[h]
\centering
\includegraphics[width=\textwidth]{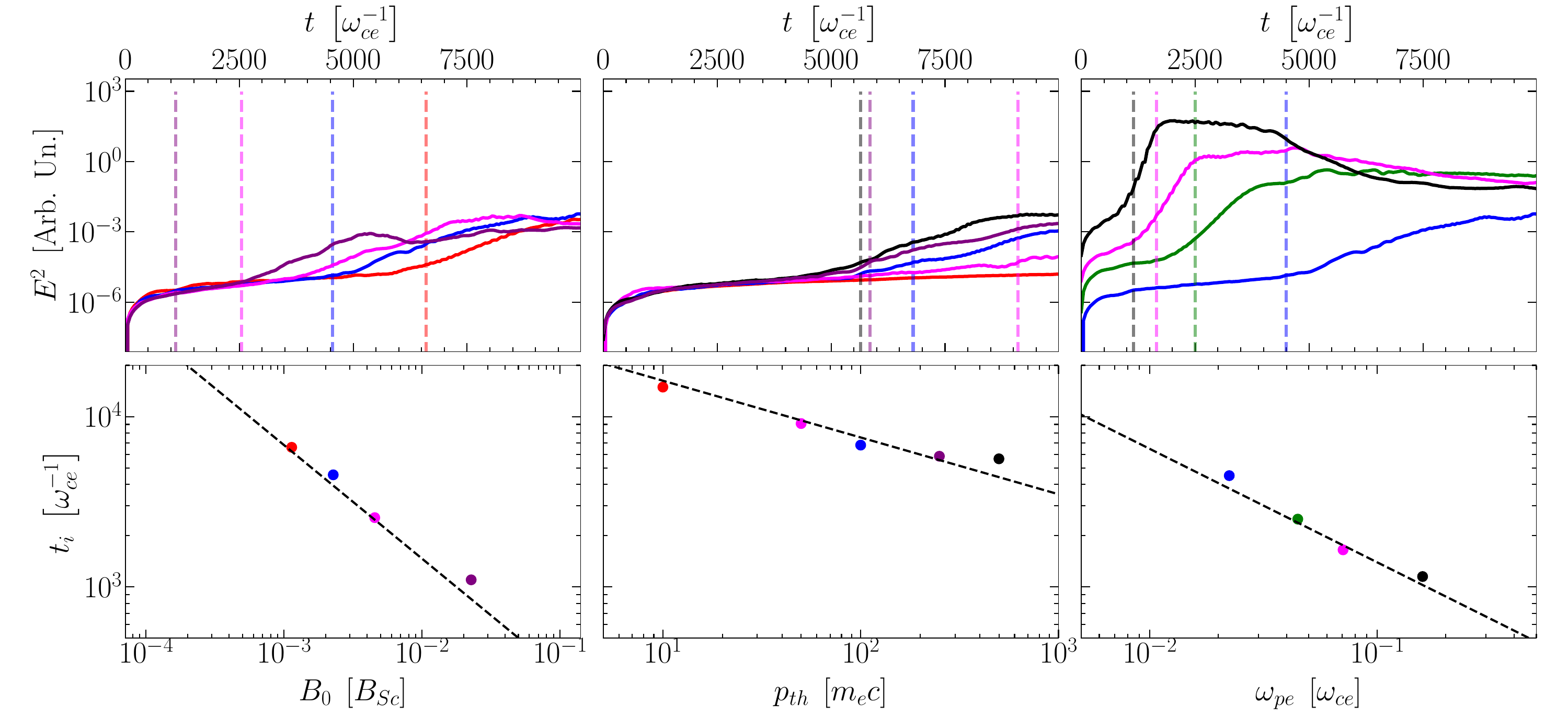}
\caption{Particle-in-cell simulation results demonstrate the correlation between the onset of electron cyclotron maser instability and the relevant parameters. The top row depicts the field energy for different simulations, each varying a single parameter per column. In the bottom row, the onset time for each simulation is presented in a log-log scale, showcasing its dependence on the varying parameter. The first column shows the results from varying the magnetic field $B_0=B/B_{Sc}$. The second column shows the results from a varying initial thermal spread $p_{th}$. The last column shows the results from varying the plasma frequency $\omega_{pe}$. The simulation outcomes are compared against the expected dependencies from Eq. (8) in the main text. Dashed lines represent the expected dependencies on $B_0$, $p_{th}$, and $\omega_{pe}$, which are $t_o\propto B_0^{-1/2}$ (for a fixed ($\omega_{pe}/\omega_{ce}$ ratio)),  and$t_o\propto p_{th}^{-1/2}\, \omega_{pe}^{-1}$, respectively.}\label{fig:sup:scaling}
\end{figure*}
In order to demonstrate the validity of the scaling of the onset time (\emph{i.e.} Eq. (\ref{eq:onset_time})) a set of PIC simulations were performed while varying the key parameters. The results of this simulation campaign (displayed in Fig. \ref{fig:sup:scaling}) confirm our understanding of the tiemscales and onset of the instability.

The simulations performed are 1D3V, with the magnetic field being aligned perpendicular to the $x_1$-direction. This allows the propagation of X-mode waves with $k$ perpendicular to $B$. The parameter scan confirms that the onset time in cyclotron periods scales as $t_o \propto p_{th}^{-1/2} \omega_{pe}^{-1}$. Moreover, an extra paramer scan was performed varying the magnetic field while keeping $\omega_{ce}/\omega_{pe}$ constant. In doing so, as $B_0$ increases so does $\omega_{pe}$. As $t_o\propto B_0^{1/2}/\omega_{pe}$, the resulting dependence is that $t_o \propto B_0^{-1/2}$ From the energy stored in the electromagnetic component $E_3$, \emph{i.e.}, the electric field associated with the X-mode. We can determine the time onset of the instability when the change in the slope of the energy over time is significant when compared with the maximum energy achieved, the fitting code checks for the condition $\Delta E^2/\Delta t>0.001 E^2_\text{max}$.

\subsection*{Timescales for collisional relaxation}
We consider three relevant collisional processes that can diffuse the ring distribution before the onset of the electron cyclotron maser instability (ECMI). These processes are: (i) Coulomb collisions, (ii) pair annihilation, and (iii) Compton scattering from synchrotron self-emission. The relaxation timescale for each process is defined as the inverse of its corresponding collision frequency.

i) The relaxation timescale due to Coulomb collisions is given by $t_{ee} =\frac{12 \pi^{3/2}}{\sqrt{2}}\frac{\epsilon_0^2 m_e^2 c^3}{e^4} \frac{ 1}{n \ln{\Lambda}}$ \cite{trubnikov1965particle, goldston1995introduction, bilbao2024ring}, where $n$ is the plasma density, $\ln{\Lambda}$ is the Coulomb logarithm, and, as we are dealing with relativistic plasmas, we have approximated $v_e \sim c$. Alternatively, $t_{ee} \simeq 5/(2 \sigma_T c n)$ and $t_{ee}\, [\mathrm{s}]\simeq1.25\times10^{14}/(n\, [\mathrm{cm}^{-3}])$, where $\sigma_T$ is the Thomson cross-section.

ii) For pair annihilation, the relaxation timescale is estimated using the cross-section for electron-positron collisions, which can be approximated by the Thomson cross-section $\sigma_T$ \cite{lightman1982relativistic, jauch2012theory}. In this case, the plasma is not simply diffused but rather "evaporates" as electron-positron pairs annihilate and are converted into high-energy photons. This process removes particles from the plasma, leading to its gradual depletion. The timescale for this "evaporation" process is $t_{eva}=1/(2\sigma_T cn)$. We note that this timescale is comparable to $t_{ee}$.

iii) The relaxation timescale due to Compton scattering, induced by synchrotron self-emission, can also be estimated. For the purpose of this discussion, an overestimate of the collisional effects suffices. The collisional frequency is defined as $\nu_{e\gamma} = 2 c \sigma_T n_\gamma$, where $n_\gamma$ is the photon density. The photon density can be estimated as the energy budget divided by the average energy per photon, \emph{i.e.} $n_\gamma = \Delta E / \hbar \left<\omega\right>$, where $\Delta E$ is the change in energy of the electron population as it cools, $\left<\omega\right>$ is the average photon angular frequency, and $\hbar$ is the reduced Planck constant. The change in energy in the electron population is $\Delta E = p_{th}^2 n_e \tau/(1+ p_{th} \tau)$, obtained from the equations of motion \cite{bilbao2022radiation,bilbao2024ring}, where $\tau = \frac{2}{3} \alpha B_o t \omega_{ce}^{-1}$ with $t$ being the time elapsed since the beginning of the cooling, and $p_{th}$ is the initial thermal spread. We assume the average frequency $\left<\omega\right>$ to be the critical frequency for a given ring radius at time $t_o$, which may underestimate the actual average frequency since electrons have higher energies earlier in the process. This overestimates the photon density at time $t$. This can be used to estimate the photon density and subsequently the relaxation time due to synchrotron self-emission at the onset time $t_o$. $t_{e\gamma} = 1/(2 \sigma_T p_\mathrm{th}^{1/3} B_0 n_e^{1/3} n_o^{2/3}) $, where $n_o =596488\, \mathrm{cm}^{-3}$, and $t_{e\gamma}\, [3.5\times10^{8}\, \mathrm{s}] = B_0^{-1} (n_e\, [\mathrm{cm}^{-3}]\ p_{th}\, [m_e c])^{-1/3}$. This estimate demonstrates that before the onset of the ECMI, the the timescale of relaxation due to compton self-synchrotron scattering is much larger than Coulomb collisions, unless the plasma is highly magnetised or the thermal energy of the distribution is much higher than the parameter space we are interested in.

In this estimate, we have neglected the non-linear Breit-Wheeler mechanism, where photons convert into pairs. This reduces the photon density, which would lengthen the relaxation timescale. Moreover, while the plasma is large, high-energy photons can have mean free paths longer than the plasma's spatial scale, allowing some photons to escape without contributing to plasma relaxation. For further details on the balance between self-absorption and emission, refer to Ochs (2024) \cite{ochs2024synchrotron}.

By comparing these timescales against the onset time, we obtain the parameter space in which the maser can operate, as shown in Fig. 5 of the main text.

\end{document}